\documentclass[aps,twocolumn]{revtex4}

\RequirePackage[T2A]{fontenc}

\usepackage{amssymb}
\usepackage{amsmath}
\usepackage{bm}
\usepackage[dvips]{epsfig}
\usepackage{graphicx}

\begin{document}

\title{On the Novikov problem for dihedral symmetry potentials}

\author{A.Ya. Maltsev}

\affiliation{
\centerline{\it L.D. Landau Institute for Theoretical Physics}
\centerline{\it 142432 Chernogolovka, pr. Ak. Semenova 1A,
maltsev@itp.ac.ru}}

\begin{abstract}
 We consider Novikov's problem of describing level lines 
of quasiperiodic functions on a plane for two-dimensional 
potentials of dihedral symmetry. It is shown that quasiperiodic 
potentials of this type can have open level lines only at 
a single energy level $\, \epsilon = \epsilon_{0} \, $, which 
brings them close to random potentials on a plane.
\end{abstract}

\maketitle

\section{Introduction}

 In this paper we consider the Novikov problem for 
quasiperiodic functions of a special symmetry, which 
is inherent, in particular, to many quasicrystals. 
In the general case, the Novikov problem consists 
in describing the level lines of quasiperiodic 
functions on a plane with an arbitrary number of 
quasiperiods. According to the general definition, 
we call a quasiperiodic function on a plane with 
$\, N \, $ quasiperiods the restriction 
of an $\, N $ - periodic function 
$\, f ({\bf z}) \, = \, f (z^{1}, \dots , z^{N} ) \, $ 
in $\, \mathbb{R}^{N} \, $ to the plane 
$\, \Pi = \mathbb{R}^{2} \, $ under a generic affine 
embedding $\, \mathbb{R}^{2} \rightarrow \mathbb{R}^{N} \, $:
\begin{equation}
\label{Vxy}
V (x, y) \,\,\, = \,\,\, f ({\bf z}) 
\Big|_{\mathbb{R}^{2} \subset \mathbb{R}^{N}}
\end{equation}

 We will call quasiperiodic functions on the plane 
potentials here and use for them the notation 
$\, V ({\bf r}) \, = \, V (x, y) \, $.

 The basis for considering Novikov's problem is 
the description of open (non-closed) level lines
$$ V (x, y) \,\,\, = \,\,\, {\rm const} $$

 The most thoroughly studied case of the Novikov 
problem is the case $\, N = 3 \, $ (see 
\cite{MultValAnMorseTheory,zorich1,dynn1992,Tsarev,
dynn1,zorich2,DynnBuDA,dynn2,dynn3}). 
As is also well known, this case plays a very important 
role in the description of galvanomagnetic phenomena 
in metals with complex Fermi surfaces (see, for example,
\cite{PismaZhETF,UFN,BullBrazMathSoc,JournStatPhys,
UMNObzor,DynMalNovUMN}). Very deep analytical results 
have been obtained by now also for the case $\, N = 4 \, $ 
(\cite{NovKvazFunc,DynNov}). A number of general results 
have also been obtained for arbitrary $\, N \, $ 
(\cite{DynMalNovUMN,BigQuas}); however, in general, 
the case $\, N > 3 \, $ has been studied in significantly 
less detail compared to $\, N = 3 \, $.

 The specificity of quasiperiodic potentials is that 
they inherit the features of both periodic and random 
potentials, representing some ``intermediate'' type 
between these two potential types. This feature also 
manifests itself in the Novikov problem, where it is 
expressed in the behavior of open level lines 
$\, V (x, y) \, = \, {\rm const} \, $. Namely, from 
the point of view of the Novikov problem, quasiperiodic 
potentials on the plane can be divided into two main types.

\begin{figure}[t]
\begin{center}
\includegraphics[width=\linewidth]{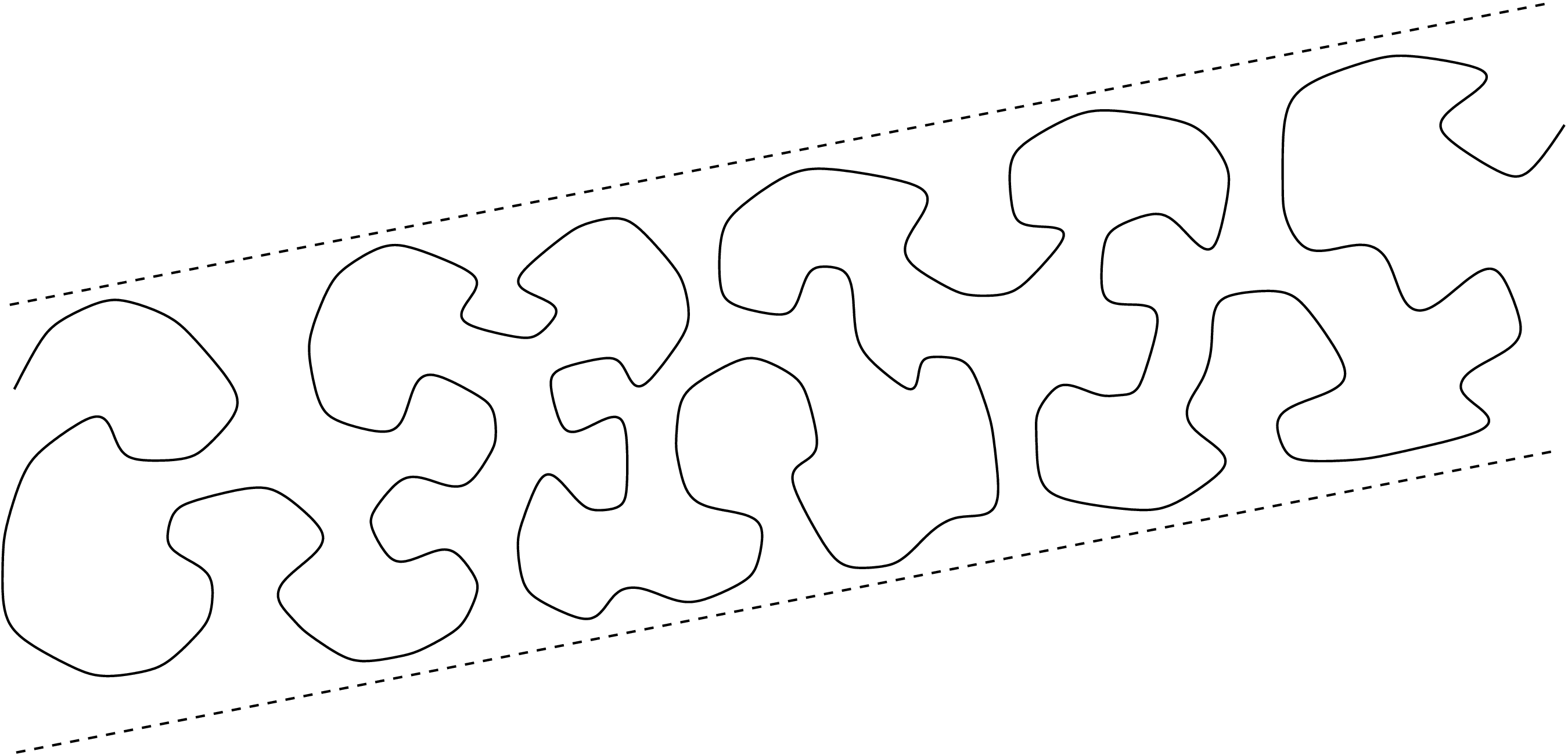}
\end{center}
\caption{General form of a ``topologically regular'' open 
level line of a quasiperiodic potential on a plane.}
\label{RegLine}
\end{figure}

  Potentials of the first type have ``topologically regular'' 
open level lines, stable with respect to small variations of 
the problem parameters. Each such level line, although not 
periodic, nevertheless lies in a straight strip of finite 
width in the plane $\, \mathbb{R}^{2} \, $, passing through it 
(Fig. \ref{RegLine}). For a given function $\, f ({\bf z}) \, $ 
such level lines arise simultaneously (at the same energy values) 
in all planes $\, \Pi \, $ of a given embedding direction 
$\, \mathbb{R}^{2} \rightarrow \mathbb{R}^{N} \, $ 
(for the restriction (\ref{Vxy})) and have the same mean 
direction in all such planes.

\begin{figure}[t]
\begin{center}
\includegraphics[width=\linewidth]{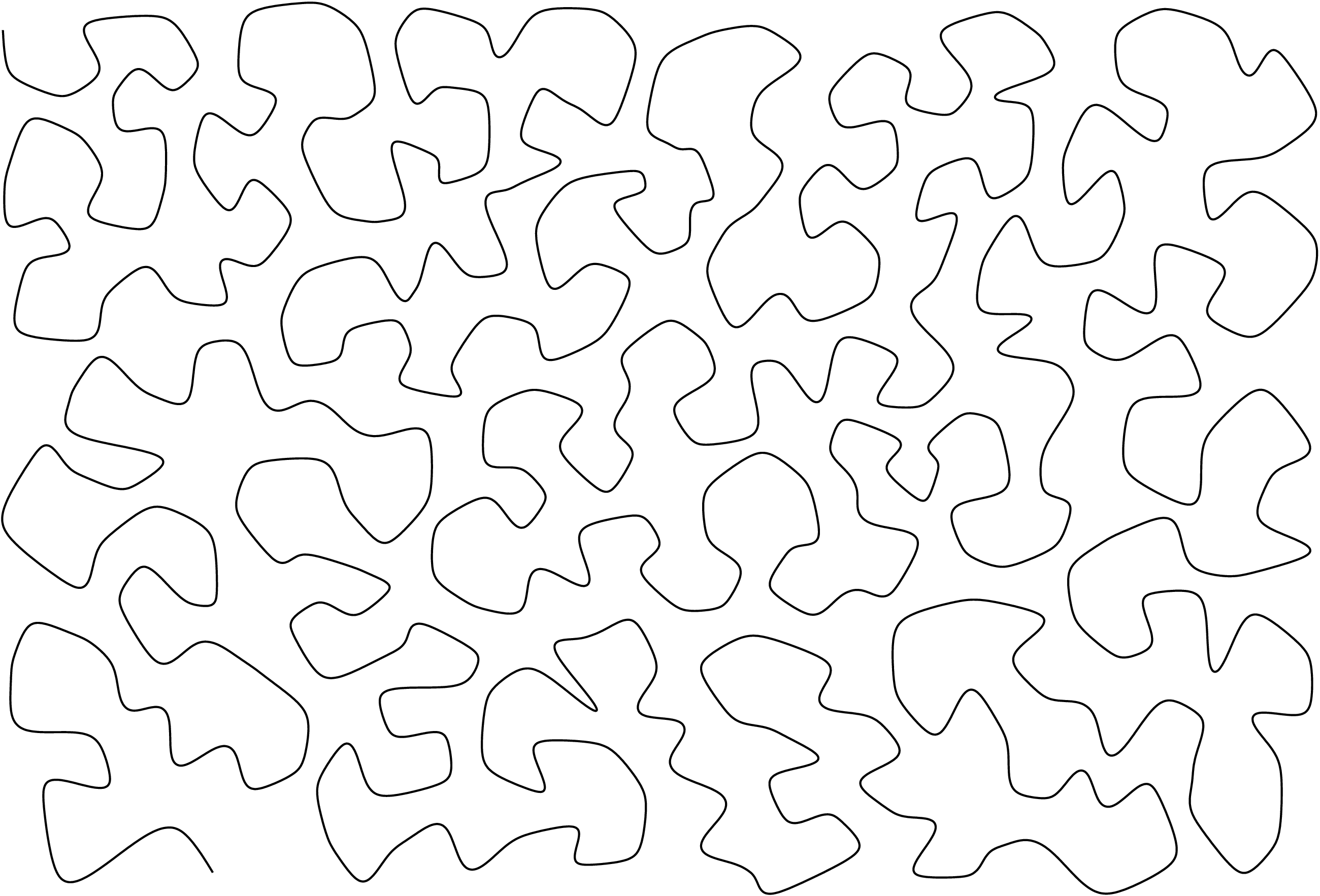}
\end{center}
\caption{General form of a ``chaotic'' level line 
of a quasiperiodic potential on a plane.}
\label{ChaoticLine}
\end{figure}

 Each stable family of ``topologically regular'' level 
lines belongs to a certain ``topological class'' determined 
by the topology of their carriers in the space 
$ \, \mathbb{R}^{N} \, $. The features of topologically 
regular level lines (as well as their topological classes) 
play an important role in describing many phenomena (see
\cite{PismaZhETF,UFN,BullBrazMathSoc,JournStatPhys,UMNObzor,
DynMalNovUMN,AnnPhys}). At the same time, potentials with 
more complex open level lines, which we will call ``chaotic'' 
here, are also of great interest. Chaotic 
(not topologically regular) level lines of quasiperiodic 
potentials usually have a more complex geometry, 
wandering ``everywhere'' on the plane $\, \mathbb{R}^{2} \, $ 
(Fig. \ref{ChaoticLine}). Potentials that have chaotic level 
lines are closer to random potentials on the plane. We will 
call such potentials ``chaotic'' here.

 Chaotic open level lines of quasiperiodic potentials 
are usually unstable to small variations of the problem 
parameters, including the energy level of the potential. 
In many interesting families of quasiperiodic potentials 
(depending on many parameters), the emergence 
of ``topologically regular'' potentials $\, V (x, y) \, $ 
corresponds to a (finite or infinite) family of regions 
in the parameter space that determine stable families of 
topologically regular open level lines. ``Chaotic'' 
potentials $\, V (x, y) \, $ arise on the complement of 
the above-described regions, resembling
the Cantor set in the parameter space.

 A very detailed study of the shape of chaotic level lines, 
as well as the sets of occurrence of ``chaotic'' potentials, 
for important families of potentials with 3 quasiperiods 
can be found in the works 
\cite{Tsarev,DynnBuDA,dynn2,Zorich1996,ZorichAMS1997,
Zorich1997,zorich3,DeLeo1,DeLeo2,DeLeo3,ZorichLesHouches,
DeLeoDynnikov1,dynn4,DeLeoDynnikov2,Skripchenko1,
Skripchenko2,DynnSkrip1,DynnSkrip2,AvilaHubSkrip1,
AvilaHubSkrip2,TrMian,DynHubSkrip}.
An important property of chaotic level lines in the case 
$\, N = 3 \, $ (\cite{dynn1}) is that they can arise only 
at a single level of the corresponding potential 
$\, V (x, y) \, $:
$$V (x, y) \,\,\, = \,\,\, \epsilon_{0} $$
(the level lines for all 
$\, \epsilon\, \neq \, \epsilon_{0} \, $ are closed). 
In addition to the complex shape of chaotic level lines, 
this property also brings ``chaotic'' potentials 
$\, V (x, y) \, $ closer to the type of random 
potentials on a plane (and also distinguishes them 
from ``topologically regular'' potentials).

 For the case $\, N > 3\, $ a similar property of 
chaotic level lines has not yet been established in 
the general case. However, there are important classes 
of quasiperiodic potentials with chaotic level lines 
for which the indicated property holds. In particular, 
in \cite{Superpos} this property was proved 
for ``two-layer'' potentials defined by the 
superposition of two periodic potentials of the same 
rotational symmetry. The reasoning in \cite{Superpos} 
is based on the approximation of quasiperiodic potentials 
by periodic potentials of the same symmetry. However, 
there are important families of quasiperiodic potentials 
for which approximation by periodic potentials of a given 
symmetry is impossible (for example, quasicrystals with 5th 
order symmetry, etc.). Due to a certain (quasi-crystalline) 
symmetry, such potentials also cannot have topologically 
regular level lines, and their open level lines must be 
classified as chaotic. Here we consider quasiperiodic 
potentials with an arbitrary $\, N \, $, possessing 
a symmetry group $\, D_{n} \, $ (a dihedral group), 
and prove the above property for them. We note here that, 
together with a complex geometry of open trajectories, 
this property also brings such potentials closer to random 
potentials on the plane.

\section{Open level lines of dihedral symmetry potentials}
\setcounter{equation}{0}

\begin{figure}[t]
\begin{center}
\includegraphics[width=\linewidth]{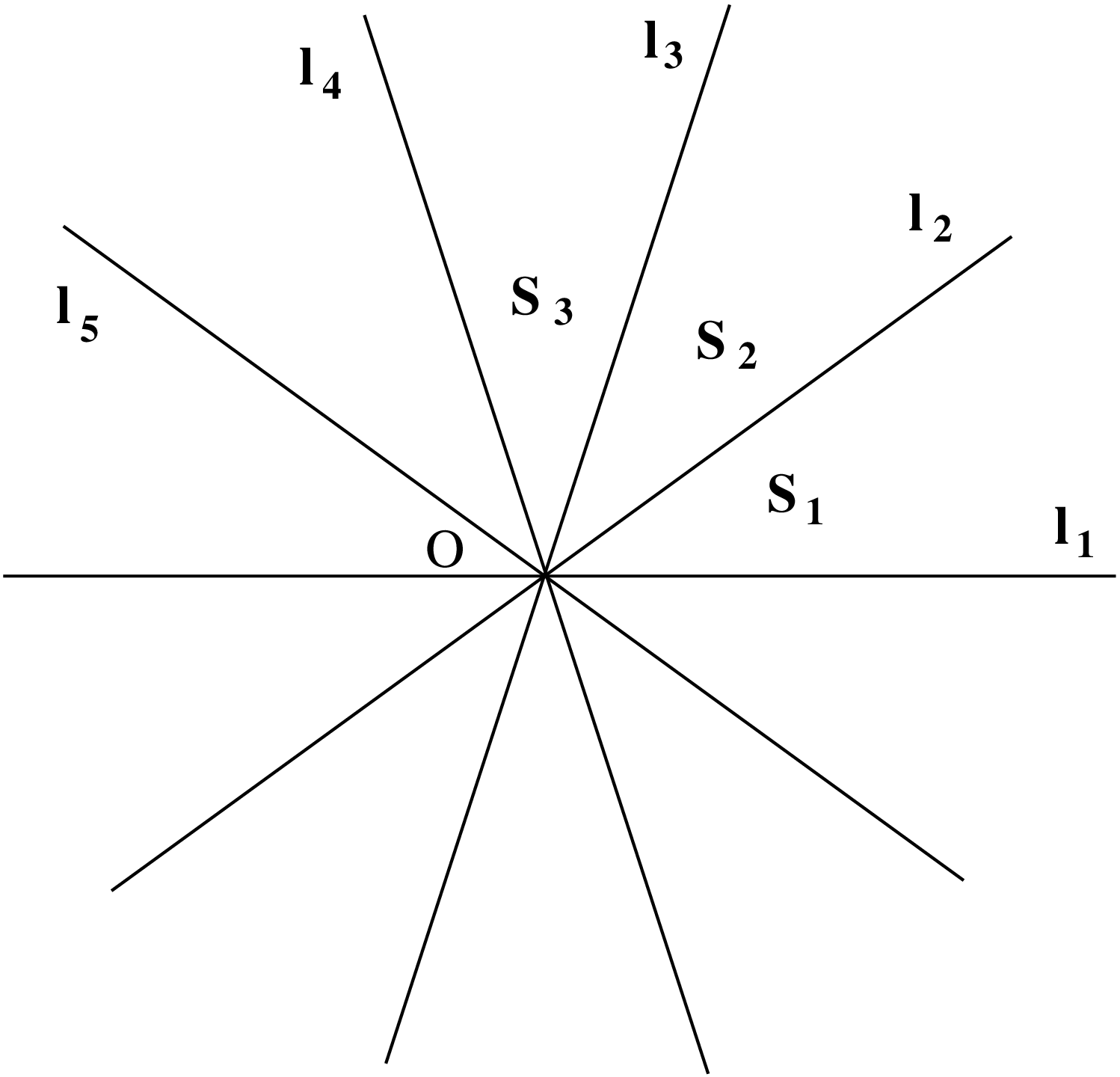}
\end{center}
\caption{Symmetry axes of the potential $\, V (x, y) \, $, 
passing through one point in the plane $\, \Pi \, $ and 
dividing the plane into $\, 2 n \, $ sectors $\, S_{i} $.}
\label{SymDiedr}
\end{figure}

 As we have already said, we will consider quasiperiodic 
potentials that have a dihedral symmetry group $\, D_{n} \, $ 
for $\, n \geq 3 \, $. This means, in particular, the existence 
of $\, n \, $ symmetry axes of the potential $\, V (x, y) \, $ 
passing through the same point $\, O \, $ on the plane 
$\, \Pi \, $ (Fig. \ref{SymDiedr}). The point $\, O \, $ is 
also the center of rotational symmetry (at angles 
$\, \alpha_{s} \, = \, 2 \pi s / n $) of the 
potential $\, V (x, y) \, $.

 The potential $\, V ({\bf r}) \, $ is quasiperiodic, 
i.e. it is determined by the restriction 
of an $\, M $ - periodic function 
$\, F ({\bf w}) \, = \, F (w^{1}, \dots , w^{M}) \, $ 
to the plane $\, \Pi \, = \, \mathbb{R}^{2} \, $ under 
some affine embedding
$\, \mathbb{R}^{2} \subset \mathbb{R}^{M} \, $:
$$V ({\bf r}) \,\,\, = \,\,\, F ({\bf w}) 
\Big|_{\Pi = \mathbb{R}^{2} \subset \mathbb{R}^{M}} $$

 In the general case, we assume that the function 
$\, F ({\bf w}) \, $ is periodic with respect to some 
lattice $\, \mathbb{Z}^{M} \, \subset \, \mathbb{R}^{M} \, $, 
which we will call integer lattice. The symmetry of the 
potential $\, V ({\bf r}) \, $ naturally corresponds to 
the symmetry of the embedding 
$\, \mathbb{R}^{2} \subset \mathbb{R}^{M} \, $, 
as well as of the lattice 
$\, \mathbb{Z}^{2} \subset \mathbb{R}^{M} \, $.

 Note here that in the examples the dimension $\, M \, $ 
is often chosen in order to make the symmetry of such an 
embedding most visual. The number of quasi-periods of the 
potential $\, V ({\bf r}) \, $ is then equal to the 
dimension of the minimal integer subspace 
$\, \mathbb{R}^{N} \subset \mathbb{R}^{M} \, $ 
containing the plane $\, \Pi \, $:
$$\Pi \,\,\, = \,\,\, \mathbb{R}^{2} \,\,\, \subset \,\,\,
\mathbb{R}^{N} \,\,\, \subset \,\,\, \mathbb{R}^{M} $$

 As an example, we can point out the embeddings 
$\, \mathbb{R}^{2} \subset \mathbb{R}^{n} \, $, 
corresponding to the $\, n $-th order symmetry, 
such that the plane $\, \Pi \, $ is contained in the 
hyperplane orthogonal to the vector 
$\, (1, 1, \dots , 1) \, $, and the corresponding 
potentials $\, V ({\bf r}) \, $ have $\, (n - 1) \, $ 
quasi-periods.
 
 In this paper we will not specify the number of 
quasi-periods of the potentials $\, V ({\bf r}) \, $. 
Here we will always assume that the potential 
$\, V ({\bf r}) \, $ is determined by the restriction 
(\ref{Vxy}) for some function 
$\, f (z^{1}, \dots , z^{N}) \, $, periodic with 
respect to some lattice 
$\, \mathbb{Z}^{N} \subset \mathbb{R}^{N} \, $,  
under some ``completely irrational'' embedding 
$\, \Pi = \mathbb{R}^{2} \subset \mathbb{R}^{N} \, $ 
(such that the plane $\, \Pi \, $ is not contained 
in any rational hyperplane and does not contain 
rational one-dimensional directions in $\, \mathbb{R}^{N}$).
 
 We will also assume here that the function 
$\, f ({\bf z}) \, $ is sufficiently smooth, and, 
in particular
\begin{equation}
\label{NablaVal}
\big| \nabla f ({\bf z}) \big| \,\,\, \leq \,\,\, C
\end{equation}
for some constant $\, C \, $.

 For simplicity, we will also assume that all 
potentials (\ref{Vxy}) can have singularities 
of only a certain type (in $\, \mathbb{R}^{2} $), 
namely, multiple saddles or isolated minima or maxima.

\vspace{1mm}

 When studying the Novikov problem, it turns out to be 
natural to consider the situation at once in all parallel 
planes $\, \Pi = \mathbb{R}^{2} \subset \mathbb{R}^{N} \, $ 
of a given direction $\, \xi \in G_{N,2} \, $, differing 
only in shifts in space $\, \mathbb{R}^{N} \, $. We will 
therefore consider here the whole family of potentials
$$V ({\bf r}, \, {\bf a}) \,\,\, , \quad \quad 
{\bf a} \, \in \, \mathbb{R}^{N - 2} \,\,\, , $$
where the vector $\, {\bf a} \, $ is orthogonal to the 
direction $\, \xi \, $, and the potential 
$\, V ({\bf r}, \, {\bf a}) \, $ is given by the 
restriction of $\, f ({\bf z}) \, $ to the plane 
$\, \Pi ({\bf a}) \, $ obtained by shifting the plane 
$\, \Pi \, $ by the vector $\, {\bf a} \, $ in 
$\, \mathbb{R}^{N} \, $. The potential 
$\, V ({\bf r}, \, {\bf 0}) \, $ coincides with the 
initial potential $\, V ({\bf r}) \, $. We also note 
that in the theory of quasicrystals the transformations 
$\, V ({\bf r}) \, \rightarrow \, V ({\bf r}, \, {\bf a}) \, $ 
are usually called phase transformations.

 Among the potentials $\, V ({\bf r}, \, {\bf a}) \, $,  
only some have the exact symmetry $\, D_{n} \, $ 
(with different positions of the point 
$\, O ({\bf a}) \in \Pi ({\bf a})$), in particular, 
the potential $\, V ({\bf r}, \, {\bf 0}) \, $ has it. 
The set of such potentials, however, corresponds 
to an everywhere dense set in the space of parameters 
$\, {\bf a} \, $. In the general case, one can also say 
that the group $\, D_{n} \, $ is the point group of the 
quasicrystallographic group of potentials 
$\, V ({\bf r}, \, {\bf a}) \, $ 
(see, for example, \cite{LePiunSad}).

 According to the general results of 
\cite{dynn3,DynMalNovUMN}, open level lines
\vspace{-2mm}
\begin{equation}
\label{VepsilonLines}
V ({\bf r}, \, {\bf a}) \,\,\, = \,\,\, \epsilon 
\end{equation}
arise (at least for one value of $\, {\bf a}$) 
in a connected closed interval
\vspace{-2mm}
$$\epsilon \,\,\, \in \,\,\, 
\left[ \epsilon_{1} , \, \epsilon_{2} \right] \,\,\, , $$
which can contract to a single point 
$\, \epsilon_{0} = \epsilon_{1} = \epsilon_{2} \, $. 
As was also shown in \cite{BigQuas}, in the case 
$\, \epsilon_{2} > \epsilon_{1} \, $ and
$$\epsilon \,\,\, \in \,\,\, 
\left( \epsilon_{1} , \, \epsilon_{2} \right) $$
open level lines (\ref{VepsilonLines}) must arise for
all values of $\, {\bf a} \in \mathbb{R}^{N - 2} \, $.

  The interval 
$\, \left[ \epsilon_{1} , \, \epsilon_{2} \right] \, $ 
is thus common to the entire family 
$\, V ({\bf r}, \, {\bf a}) \, $. In particular, 
the possibility of the emergence of open level lines 
(\ref{VepsilonLines}) at no more than one level 
$\, \epsilon_{0} \, $ for one of the potentials 
$\, V ({\bf r}, \, {\bf a}) \, $ implies the same 
property for all potentials of this family at once.

 Here, as we have already said, we assume that the 
potential $\, V ({\bf r}, \, {\bf 0}) \, $ has an exact 
symmetry $\, D_{n} \, $, and our task is to prove the 
relation $\, \epsilon_{1} = \epsilon_{2} = \epsilon_{0} \, $ 
in this situation. To prove it for all potentials of the 
family $\, V ({\bf r}, \, {\bf a}) \, $ it suffices 
to prove it only for the potential 
$\, V ({\bf r}, \, {\bf 0}) \, $.

\vspace{2mm}

 Let us assume that the potential 
$\, V ({\bf r}, \, {\bf 0}) \, $ has open level lines 
in some finite energy interval 
$\, \epsilon \in ( \epsilon_{1} , \, \epsilon_{2} ) \, $. 
Without loss of generality, let us choose two values 
$\, E_{1} \, $, $\, E_{2} \, $:
$$\epsilon_{1} \,\,\, < \,\,\, E_{1} \,\,\, < \,\,\,
E_{2} \,\,\, < \,\,\, \epsilon_{2} \,\,\, , $$
such that the open level lines
\begin{equation}
\label{E12Lines}
V ({\bf r}, \, {\bf 0}) \,\,\, = \,\,\, E_{1,2} 
\end{equation}
are nonsingular smooth curves.

 We assume that the potential 
$\, V ({\bf r}, \, {\bf 0}) \, $ has $\, n \, $ symmetry axes 
$\, {\bf l}_{1} \, $, $\, \dots \, $, $\, {\bf l}_{n} \, $ 
passing through the point $\, O \in \Pi \, $ and dividing the 
plane $\, \Pi = \mathbb{R}^{2} \, $ into $\, 2 n \, $ 
identical sectors 
$\, S_{1} \, $, $\, \dots \, $, $\, S_{2n} \, $ 
(Fig. \ref{SymDiedr}).

\begin{figure}[t]
\begin{center}
\includegraphics[width=\linewidth]{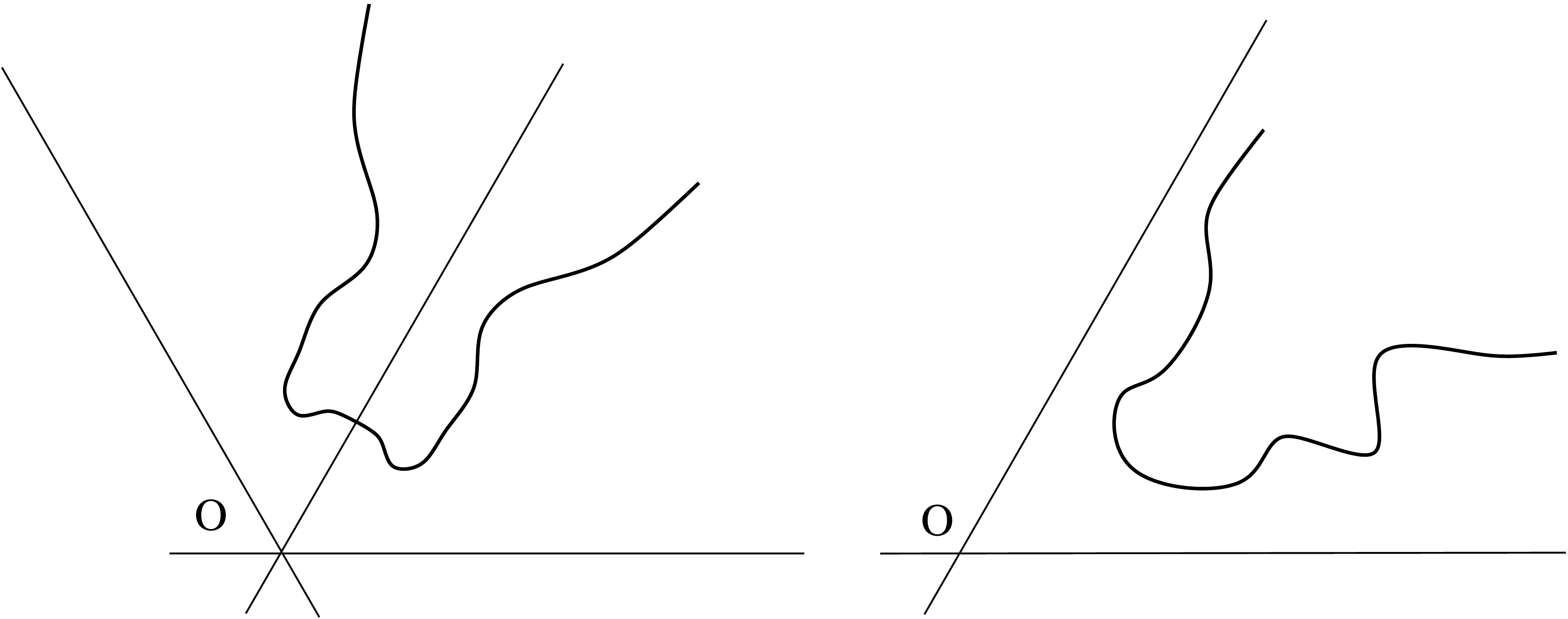}
\end{center}
\caption{Possible form of open level lines 
(\ref{E12Lines}) (schematically).}
\label{OpenLinesE12}
\end{figure}

 An open non-singular level line (\ref{E12Lines}) 
cannot intersect both rays bounding any of the 
sectors $\, S_{i} \, $ at once (otherwise, by 
reflection symmetry, it must be a closed level 
line going around the point $\, O $). Thus, any of 
the open level lines (\ref{E12Lines}) can either 
intersect one of the rays at a single point 
(and be symmetric with respect to it) or lie 
entirely inside one of the sectors 
(Fig. \ref{OpenLinesE12}).

\begin{figure}[t]
\begin{center}
\includegraphics[width=\linewidth]{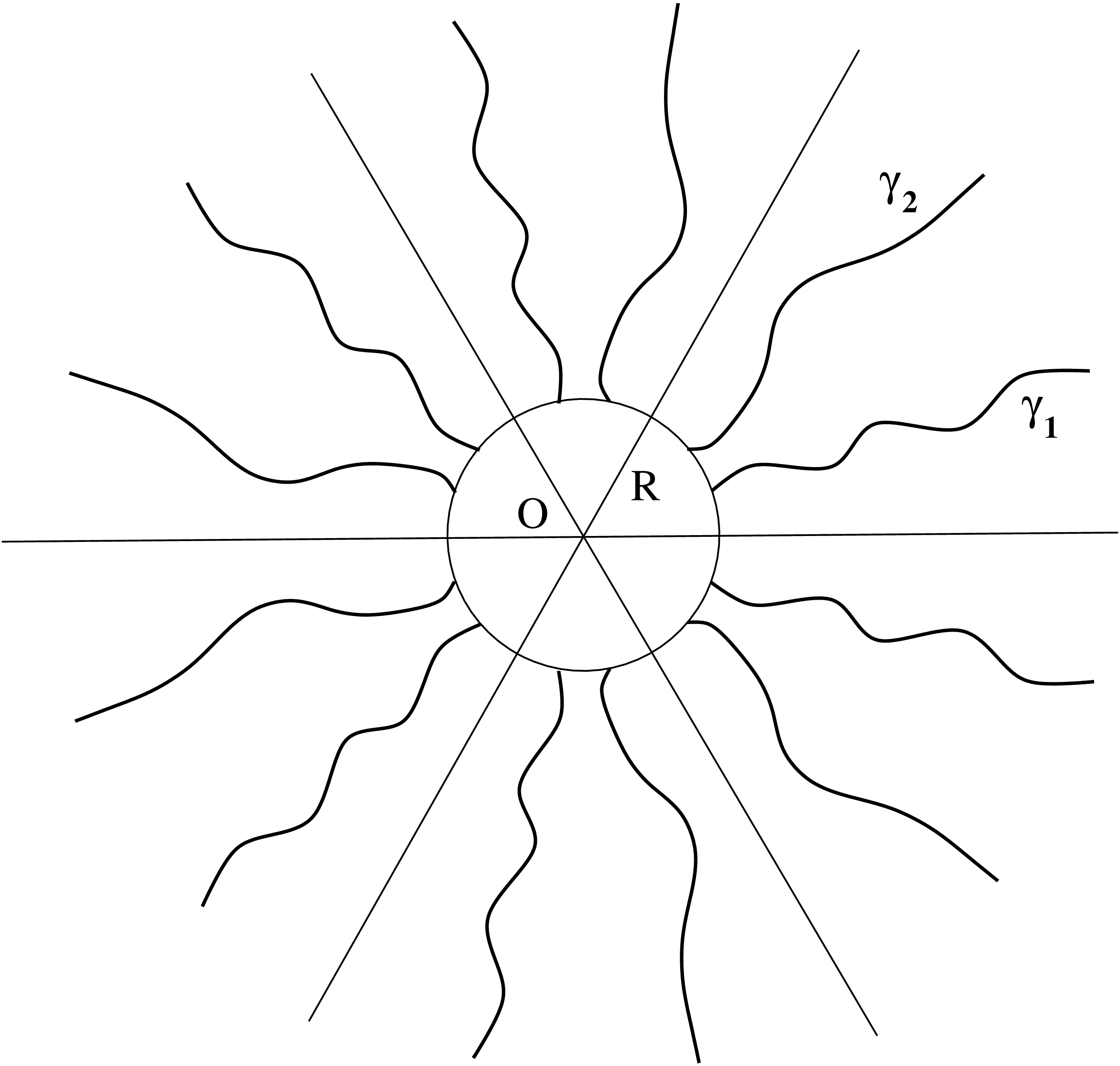}
\end{center}
\caption{Curves $\, \gamma_{1} \, $ and 
$\, \gamma_{2} \, $ lying at levels 
$\, \epsilon = E_{1} \, $ and $\, \epsilon = E_{2} \, $ 
(in each of the sectors $\, S_{i} \, $).}
\label{gamma12}
\end{figure}

 In any case, each of such level lines contains a curve 
$\, \gamma \, $ starting at some distance $\, R \, $ 
from the point $\, O \, $ and going to infinity, being 
entirely in one of the sectors $\, S_{i} \, $ and 
at a distance of at least $\, R \, $ from the point 
$\, O \, $. By symmetry, the same curves are present 
in each of the sectors $\, S_{i} \, $. Thus, we can 
indicate in each $\, S_{i} \, $ the similar curves 
$\, \gamma_{1} \, $ and $\, \gamma_{2} \, $ lying at the 
levels $\, E_{1} \, $ and $\, E_{2} \, $, respectively:
$$V ({\bf r}, \, {\bf 0}) \Big|_{\gamma_{1}} 
\,\,\, \equiv \,\,\, E_{1} \,\,\, , \quad \quad
V ({\bf r}, \, {\bf 0}) \Big|_{\gamma_{2}} 
\,\,\, \equiv \,\,\, E_{2} $$
(Fig. \ref{gamma12}).

 As we have already noted, potentials 
$\, V ({\bf r}, \, {\bf a}) \, $, possessing exact 
symmetry $\, D_{n} \, $, correspond to an everywhere 
dense set in the space of parameters $\, {\bf a} \, $. 
Such potentials, in particular, correspond to planes 
$\, \Pi ({\bf a}) \subset \mathbb{R}^{N} \, $, 
obtained by integer shifts of the original plane 
$\, \Pi = \Pi ({\bf 0}) \, $ in the space 
$\, \mathbb{R}^{N} \, $. Thus, all planes of direction 
$\, \xi \, $ passing through integer shifts of the 
point $\, O \, $ in $\, \mathbb{R}^{N} \, $ correspond 
to potentials $\, V ({\bf r}, \, {\bf a}) \, $ possessing 
exact symmetry $\, D_{n} \, $. Obviously, the picture of 
level lines in such planes is identical to the picture 
in the plane $\, \Pi ({\bf 0}) \, $, and the corresponding 
shift of the point $\, O \, $ ($O ({\bf a}) \in \Pi ({\bf a})$) 
plays in them the same role as the point $\, O \, $ in the 
plane $\, \Pi ({\bf 0}) \, $.

\begin{figure}[t]
\begin{center}
\includegraphics[width=0.9\linewidth]{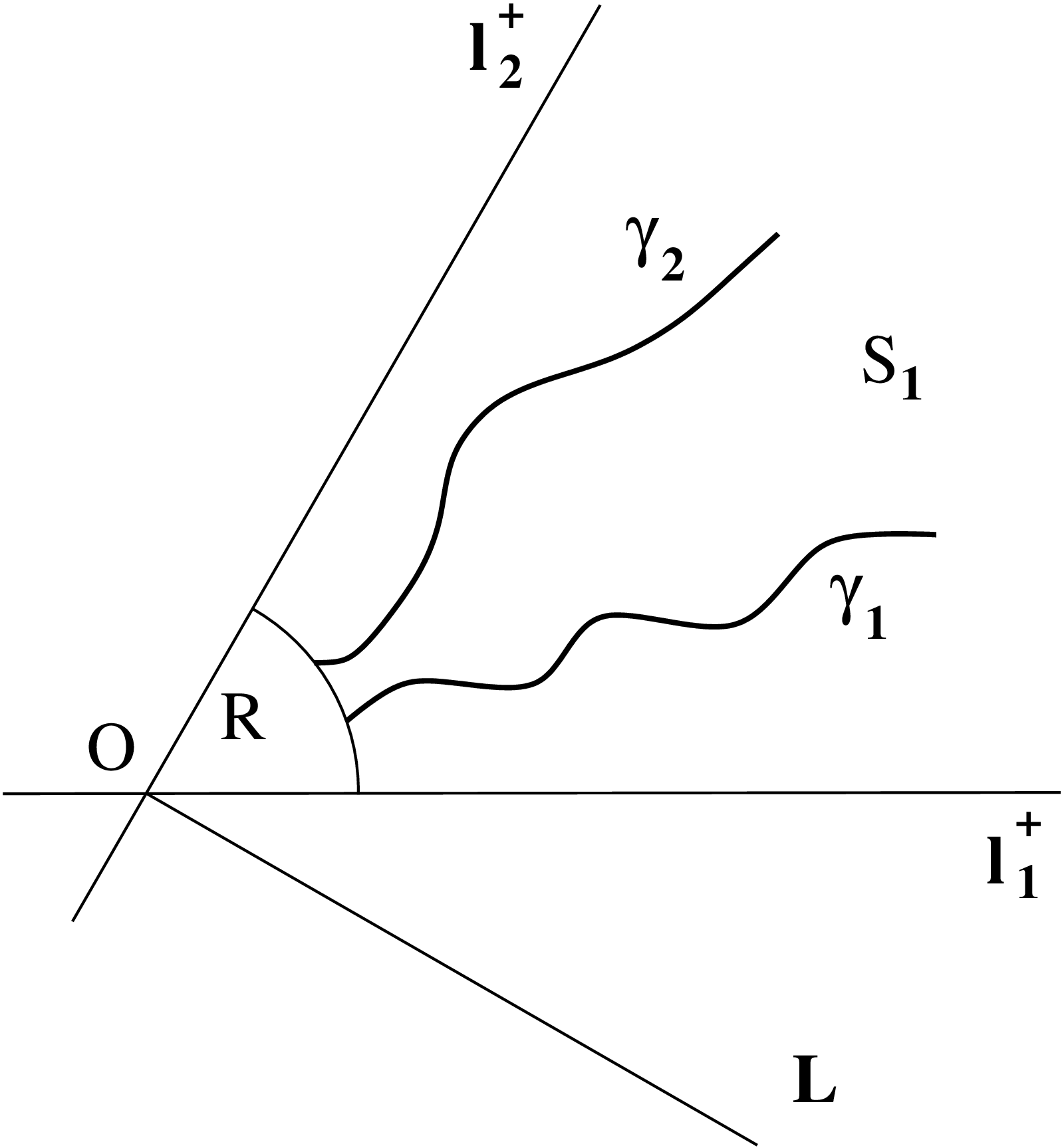}
\end{center}
\caption{Sector $\, S_{1} \, $ in the plane 
$\, \Pi ({\bf 0}) \, $ and ray $\, {\bf L} \, $, 
orthogonal to $\, {\bf l}_{2} \, $.}
\label{Sector1}
\end{figure}

 Consider in the plane $\, \Pi ({\bf 0}) \, $ the sector 
$\, S_{1} \, $, bounded by the rays $\, {\bf l}^{+}_{1} \, $ 
and $\, {\bf l}^{+}_{2} \, $, as well as the ray 
$\, {\bf L} \, $, emanating from the point $\, O \, $ 
and orthogonal to the line $\, {\bf l}_{2} \, $ 
(Fig. \ref{Sector1}).

 The ray $\, {\bf L} \subset \mathbb{R}^{N} \, $ 
forms an everywhere dense winding of the torus 
$\, \mathbb{T}^{N} \, $ after the factorization
$$\mathbb{R}^{N} \,\,\, \rightarrow \,\,\, \mathbb{T}^{N}
\,\,\, = \,\,\, \mathbb{R}^{N} \Big/ \, \mathbb{Z}^{N} $$

 This means, in particular, that $\, {\bf L} \, $ comes 
(infinitely many times) arbitrarily close to integer 
shifts of the point $\, O \, $ in $\, \mathbb{R}^{N} \, $. 
We can thus choose an integer shift of the point 
$\, O \, $ ($O^{\prime} \in \mathbb{R}^{N}$) located 
at a distance
\begin{equation}
\label{delta}
\delta \,\,\, < \,\,\, (E_{2} - E_{1}) / C 
\end{equation}
from the ray $\, {\bf L} \, $ and at a distance
$$D \,\,\, > \,\,\, 2 R \,\, + \,\, \delta $$
from the point $\, O \, $ in $\, \mathbb{R}^{N} \, $.

\begin{figure}[t]
\begin{center}
\includegraphics[width=0.9\linewidth]{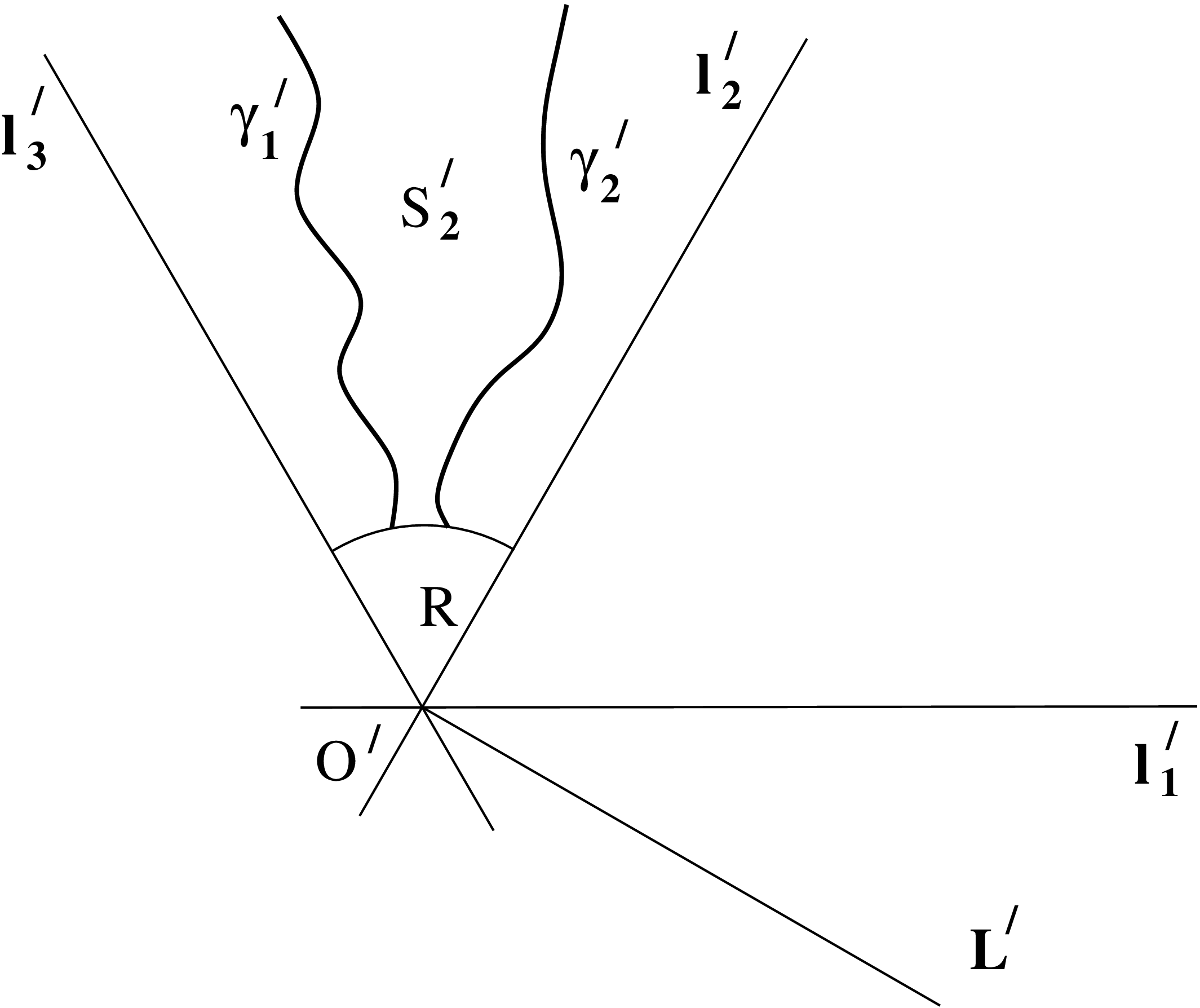}
\end{center}
\caption{Sector $\, S_{2}^{\prime} \, $ in the plane 
$\, \Pi ({\bf a}) \ni O^{\prime} \, $, containing curves 
$\, \gamma_{1}^{\prime} \, $ and 
$\, \gamma_{2}^{\prime} \, $.}
\label{Sector2}
\end{figure}

 As we have already said, the picture of level lines 
(\ref{VepsilonLines}) in the plane 
$\, \Pi ({\bf a}) \ni O^{\prime} \, $ is identical to 
the picture in the plane $\, \Pi ({\bf 0}) \, $. 
Let us consider in the plane $\, \Pi ({\bf a}) \, $ 
the sector $\, S_{2}^{\prime} \, $ and the curves 
$\, \gamma_{1}^{\prime} \, $ and 
$\, \gamma_{2}^{\prime} \, $ located in it, which are 
identical to the curves $\, \gamma_{1} \, $ and 
$\, \gamma_{2} \, $ in the sector $\, S_{2} \, $ in 
the plane $\, \Pi ({\bf 0}) \, $ (Fig. \ref{Sector2}).

 By definition, we have on the curves 
$\, \gamma_{1}^{\prime} \, $ and $\, \gamma_{2}^{\prime} \, $:
\begin{equation}
\label{gammaprimerel}
V ({\bf r}, \, {\bf a}) \Big|_{\gamma_{1}^{\prime}} 
\,\,\, \equiv \,\,\, E_{1} \,\,\, , \quad \quad
V ({\bf r}, \, {\bf a}) \Big|_{\gamma_{2}^{\prime}} 
\,\,\, \equiv \,\,\, E_{2}
\end{equation}

 Let $\, {\bf h} \, $ be the minimal vector in 
$\, \mathbb{R}^{N} \, $ connecting the point 
$\, O^{\prime} \, $ with the ray 
$\, {\bf L} \, $ ($| {\bf h} | = \delta $).

 Let $\, \sigma_{\bf h} \left[ \Pi ({\bf a}) \right] \, $ 
represent the shift of the plane $\, \Pi ({\bf a}) \, $ 
by the vector $\, {\bf h} \, $. Obviously
$$\sigma_{\bf h} \left[ \Pi ({\bf a}) \right] \,\,\, = \,\,\,
\Pi ({\bf 0}) $$

 Considering the shift
$$ \sigma_{\bf h} \left[ S_{2}^{\prime} \right] 
\,\,\, \subset \,\,\, \Pi ({\bf 0}) \,\,\, , $$
we can see that two situations are possible:

\vspace{1cm}

\noindent
A) Either $\, \gamma_{1} \, $ or $\, \gamma_{2} \, $ intersects 
$\, \sigma_{\bf h} \left[ \, {\bf l}_{2}^{\prime} \, \right] \, $.

\vspace{1mm}

\begin{figure}[t]
\begin{center}
\includegraphics[width=\linewidth]{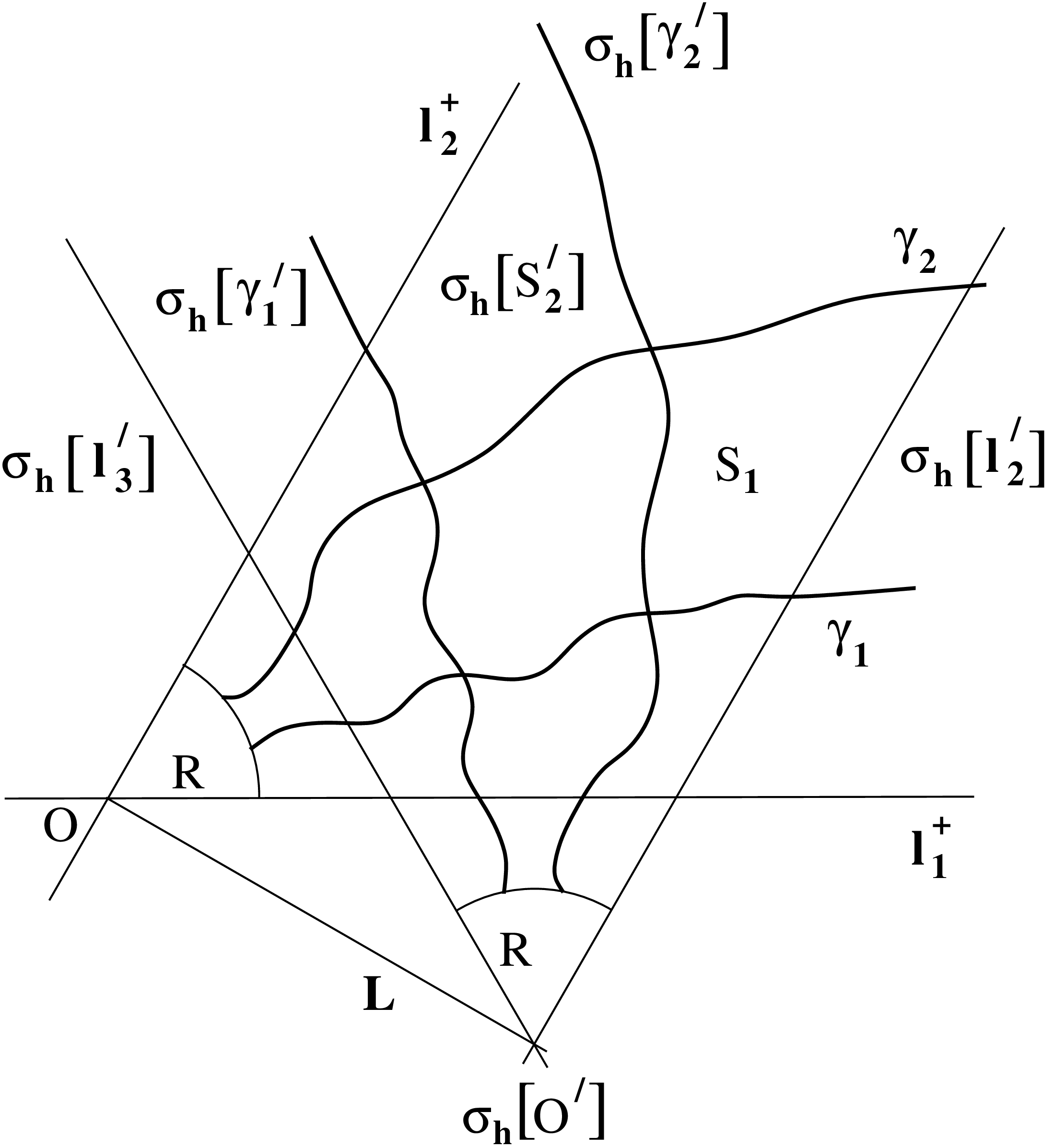}
\end{center}
\caption{Intersection of the curves 
$\, \gamma_{1} \, $, $\, \gamma_{2} \, $ with the curves 
$\, \sigma_{\bf h} \left[ \, \gamma_{2}^{\prime} \, \right] \, $, 
$\, \sigma_{\bf h} \left[ \, \gamma_{1}^{\prime} \, \right] \, $ 
in situation A.}
\label{S1S2A}
\end{figure}

 In this case, either $\, \gamma_{1} \, $ intersects 
$\, \sigma_{\bf h} \left[ \, \gamma_{2}^{\prime} \, \right] \, $ 
or $\, \gamma_{2} \, $ intersects 
$\, \sigma_{\bf h} \left[ \, \gamma_{1}^{\prime} \, \right] \, $ 
(Fig. \ref{S1S2A}).

 According to (\ref{NablaVal}), (\ref{delta}) and 
(\ref{gammaprimerel}) we then have, for example, 
in the first case
\begin{multline*}
V ({\bf r}, \, {\bf 0}) 
\Big|_{\sigma_{\bf h} [\gamma_{2}^{\prime}]} 
\,\,\, \equiv \,\,\, f ({\bf z})
\Big|_{\sigma_{\bf h} [\gamma_{2}^{\prime}]} 
\,\,\, \geq \,\,\,\,\,  
f ({\bf z}) \Big|_{\gamma_{2}^{\prime}} \,\, - \,\,
\delta \cdot C \,\,\, >  \\
> \,\,\, V ({\bf r}, \, {\bf a}) 
\Big|_{\gamma_{2}^{\prime}} \,\, - \,\, 
\left( E_{2} \, - \, E_{1} \right) \,\,\, = \,\,\, E_{1}
\quad \quad
\end{multline*}

 At the same time, we have
$$V ({\bf r}, \, {\bf 0}) \Big|_{\gamma_{1}} 
\,\,\, \equiv \,\,\, E_{1} \,\,\, , $$
which leads to a contradiction (similarly for the second case).

\vspace{1cm}

\noindent
B) The curves $\, \gamma_{1} \, $ and $\, \gamma_{2} \, $ 
do not intersect 
$\, \sigma_{\bf h} \left[ \, {\bf l}_{2}^{\prime} \, \right] \, $.

\vspace{1mm}

\begin{figure}[t]
\begin{center}
\includegraphics[width=0.9\linewidth]{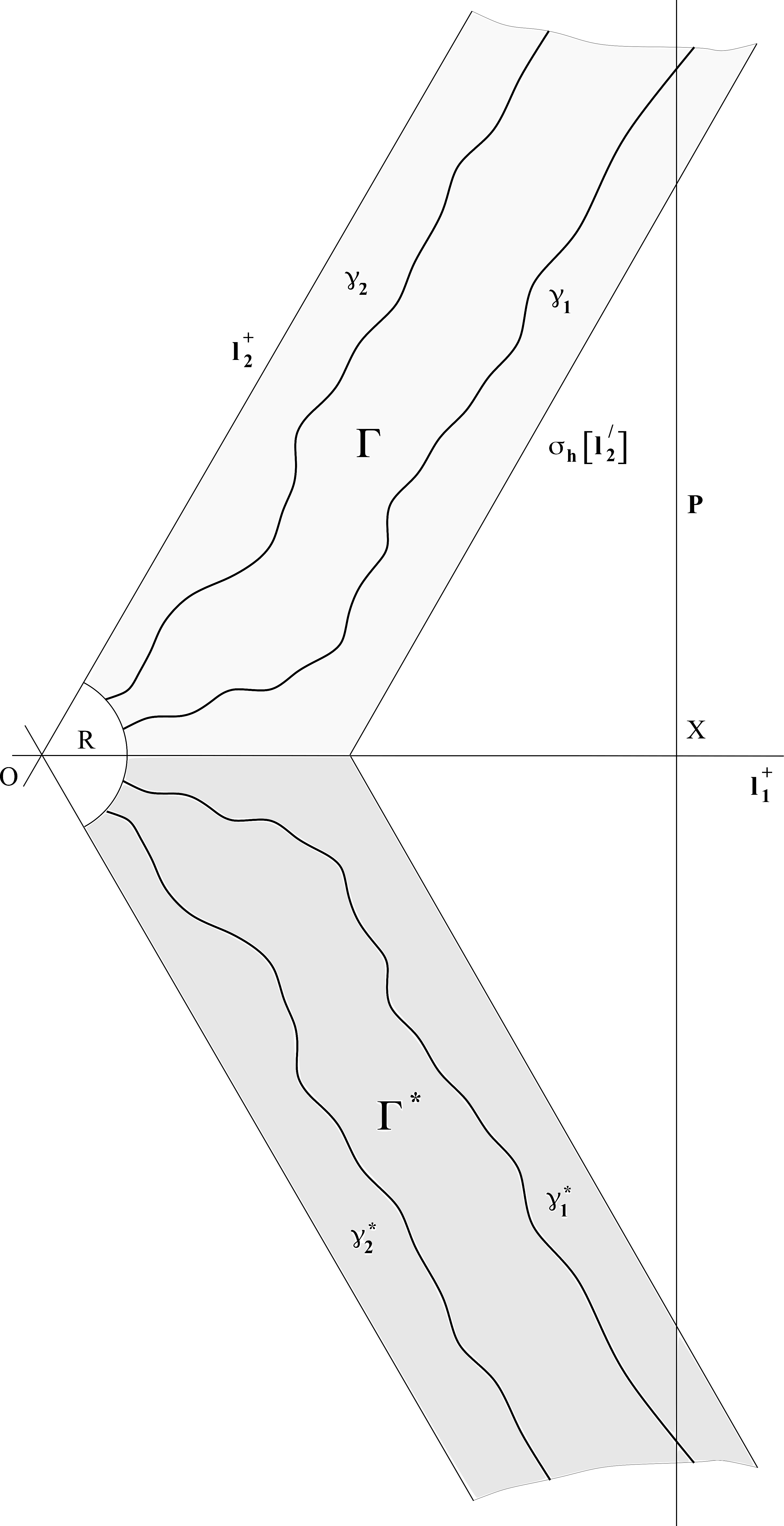}
\end{center}
\caption{Strips $\, \Gamma \, $ and $\, \Gamma^{*} \, $ 
in the plane $\, \Pi ({\bf 0}) \, $ containing curves 
$\, \gamma_{1} \, $, $\, \gamma_{2} \, $, 
$\, \gamma^{*}_{1} \, $, $\, \gamma^{*}_{2} \, $ 
in situation B.}
\label{GammaGammaStar}
\end{figure}

 In this case, both curves $\, \gamma_{1} \, $ and 
$\, \gamma_{2} \, $ lie in the ``half-strip'' $\, \Gamma \, $, 
the width of which does not exceed $\, D + \delta \, $. 
By symmetry, the sector $\, S_{2n} \, $ contains a similar 
half-strip $\, \Gamma^{*} \, $ with the curves 
$\, \gamma^{*}_{1} \, $ and $\, \gamma^{*}_{2} \, $, 
symmetrical to $\, \gamma_{1} \, $ and $\, \gamma_{2} \, $ 
with respect to $\, {\bf l}_{1} \, $ (Fig. \ref{GammaGammaStar}).

 Consider a straight line $\, {\bf P} \, $ orthogonal to 
$\, {\bf l}_{1} \, $ and lying at a sufficient distance 
(to the right) from the intersection of the strips 
$\, \Gamma \, $ and $\, \Gamma^{*} \, $ 
(Fig. \ref{GammaGammaStar}). Let 
$\, X \in \Pi ({\bf 0}) \, $ be the intersection point of 
$\, {\bf P} \, $ and $\, {\bf l}_{1} \, $.

 The line $\, {\bf P} \, $ forms a dense winding everywhere in
$$\mathbb{T}^{N} \,\,\, = \,\,\, 
\mathbb{R}^{N} \Big/ \, \mathbb{Z}^{N} $$
after factorization. We can therefore find an integer shift 
of a point $\, X \, $ ($\overline{X} \in \mathbb{R}^{N}$) 
located at a distance of at most $\, \delta \, $ from the 
line $\, {\bf P} \subset \mathbb{R}^{N} \, $ and 
at a sufficiently large distance from $\, X \, $.

 The picture of level lines in the plane 
$\, \Pi ({\bf a}) \ni \overline{X} \, $ is identical to the 
picture in the plane $\, \Pi ({\bf 0}) \, $. In particular, 
we can indicate here the stripes $\, \overline{\Gamma} \, $, 
$\, \overline{\Gamma}^{*} \, $, as well as the curves 
$\, \overline{\gamma}_{1} \, $, $\, \overline{\gamma}_{2} \, $, 
$\, \overline{\gamma}^{*}_{1} \, $, 
$\, \overline{\gamma}^{*}_{2} \, $, identical to those 
in the plane $\, \Pi ({\bf 0}) \, $.

\vspace{1mm}

\begin{figure}[t]
\begin{center}
\includegraphics[width=\linewidth]{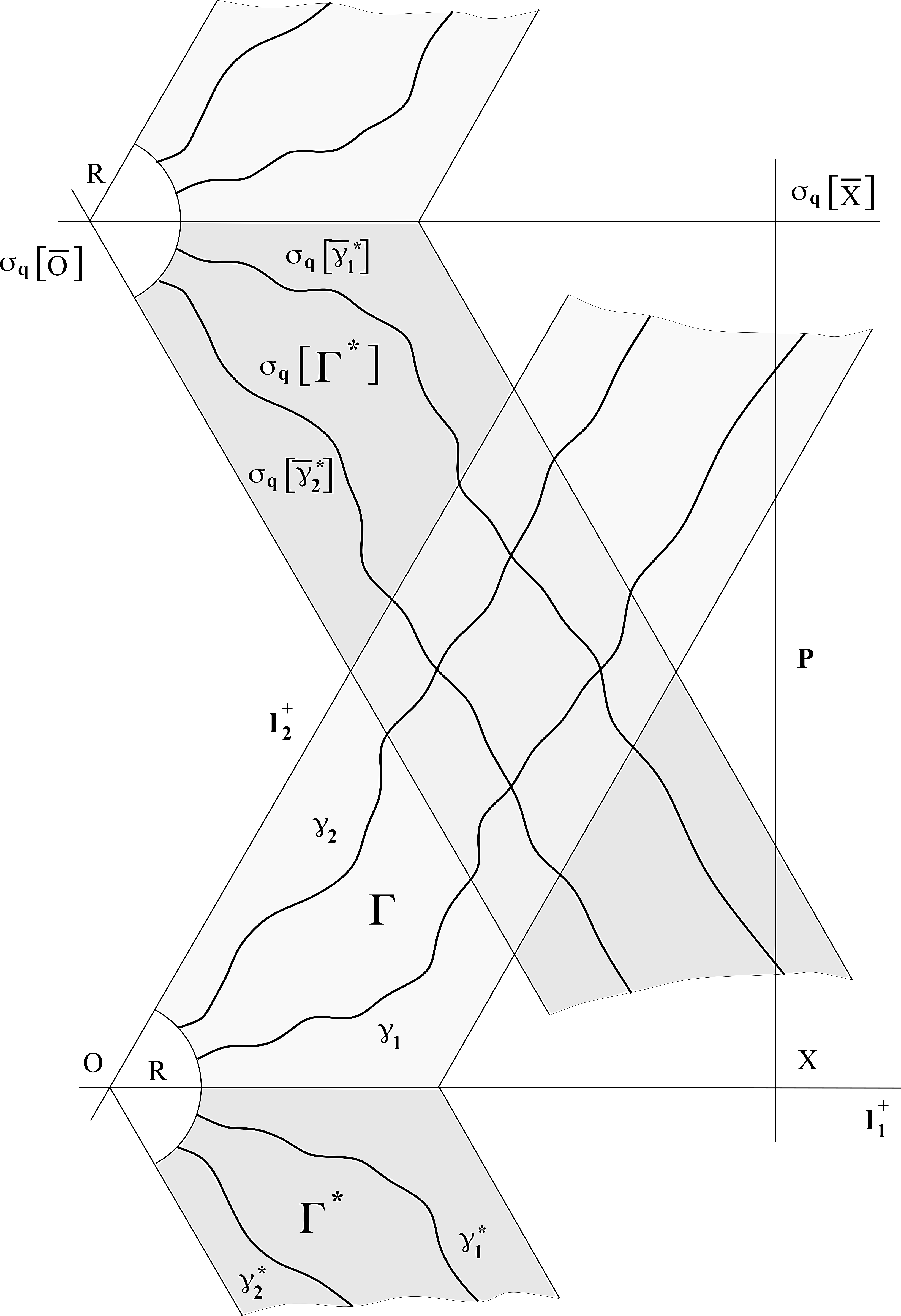}
\end{center}
\caption{Intersection of the curves 
$\, \sigma_{\bf q} \left[ \overline{\gamma}^{*}_{2} \right] \, $, 
$\, \sigma_{\bf q} \left[ \overline{\gamma}^{*}_{1} \right] \, $ 
with the curves $\, \gamma_{1} \, $ and $\, \gamma_{2} \, $ 
in the plane $\, \Pi ({\bf 0}) \, $.}
\label{Intersection}
\end{figure}

  Let $\, {\bf q} \, $ be the shortest vector connecting 
$\, \overline{X} \, $ with the line 
$\, {\bf P} \, $ ($| {\bf q} | \leq \delta $). It is easy 
to see that with a suitable choice of the point 
$\, \overline{X} \, $ (in accordance with the above 
prescription) we can ensure the intersection of the curves 
$\, \sigma_{\bf q} \left[ \overline{\gamma}^{*}_{2} \right] \, $ 
and 
$\, \sigma_{\bf q} \left[ \overline{\gamma}^{*}_{1} \right] \, $ 
with the curves $\, \gamma_{1} \, $ and $\, \gamma_{2} \, $ 
in the plane $\, \Pi ({\bf 0}) \, $ (Fig. \ref{Intersection}).

 As a result, we arrive at the same contradiction 
as in situation A.

\vspace{1cm}

 We have thus proved that under the conditions listed above, 
the emergence of open level lines for any potential 
$\, V ({\bf r}, \, {\bf a}) \, $ of the above family  
is possible only at a single energy value 
$\, \epsilon = \epsilon_{0} \, $. According to the general 
results of \cite{DynMalNovUMN}, each specific potential 
$\, V ({\bf r}, \, {\bf a}) \, $ from this family must have 
either open level lines or closed level lines of arbitrarily 
large sizes (or both) at the level 
$\, \epsilon = \epsilon_{0} \, $.

 All level lines of potentials $\, V ({\bf r}, \, {\bf a}) \, $ 
at $\, \epsilon \neq \epsilon_{0} \, $ are closed. The sizes of 
closed level lines of all $\, V ({\bf r}, \, {\bf a}) \, $ at 
$\, \epsilon \neq \epsilon_{0} \, $ are limited by one constant 
$\, D (\epsilon) \, $, which tends to infinity at 
$\, \epsilon \rightarrow \epsilon_{0} \, $. It should be said 
that the growth rate of $\, D (\epsilon) \, $ in this limit 
depends significantly on the features of the initial potential 
$\, V ({\bf r}) \, $ (and, in particular, on the number of 
quasi-periods $\, N$).

\section{Conclusion}
\setcounter{equation}{0}

 In this paper we consider the Novikov problem, i.e. the 
problem of describing the level lines of quasiperiodic 
functions on a plane, for quasiperiodic potentials admitting 
the dihedral symmetry $\, D_{n} \, $. Potentials of this 
type can have only ``chaotic'' open level lines according 
to the general terminology dividing the open level lines of 
quasiperiodic functions into topologically regular and chaotic. 
Here we show that for an arbitrary number of quasiperiods, 
such potentials can have open level lines only at a single 
energy value $\, \epsilon = \epsilon_{0} \, $. Such behavior 
of the open level lines, together with their complex geometry, 
brings these potentials close to random potentials on a plane, 
at the same time, such potentials also play an important 
role in the theory of two-dimensional quasicrystals. 
It can be noted that our results can also be useful 
for clarifying the connection between the geometry of open 
level lines and the features of their emergence in the energy 
interval in the general formulation of the Novikov problem.

\end{document}